\documentclass[letterpaper, 10 pt, conference]{ieeeconf}
\IEEEoverridecommandlockouts
\usepackage{cite}
\usepackage{amsmath,amssymb,amsfonts}
\usepackage{algorithmic}
\usepackage{graphicx}

\usepackage{graphicx}
\usepackage{subfigure}
\usepackage{epsfig}
\usepackage{cite}

\usepackage[hidelinks]{hyperref}

\usepackage[dvipsnames]{xcolor}
\usepackage{tikz}
\usepgflibrary{fpu}
\usetikzlibrary{positioning,calc,arrows,shapes,shadows,matrix,backgrounds}
\usetikzlibrary{external}

\usetikzlibrary{positioning,calc,arrows,shapes,shadows}
\tikzset{
	auto,
	node/.style={
		circle,
		draw,
		fill=gray!25,
		minimum size = \dh,
		inner sep=\dn},
	sum/.style={circle,draw,draw=black,inner sep=0mm,minimum size=2mm},
	jun/.style={circle,draw,draw=black,inner sep=0mm,minimum size=0mm},
	>={latex},
}
\def\dn{1ex*0.75}

\def\dh{6*\dn}

\usepackage{pgfplots}
\pgfplotsset{compat=1.18}

\usepackage{mathtools}

\newcommand{\set}[1]{\mathbb{#1}}
\newcommand{\N}{\set{N}}
\newcommand{\R}{\set{R}}

\newcommand{\SI}{\set{I}}
\newcommand{\SU}{\set{U}}
\newcommand{\SX}{\set{X}}
\newcommand{\SY}{\set{Y}}

\newcommand{\CK}{\mathcal K}
\newcommand{\CL}{\mathcal L}

\newcommand{\lb}{\lambda}
\newcommand{\al}{\alpha}
\newcommand{\be}{\beta}
\newcommand{\ga}{\gamma}
\newcommand{\de}{\delta}
\newcommand{\eps}{\varepsilon}

\newcommand{\abs}[1]{\left| #1 \right|}
\newcommand{\floor}[1]{\lfloor #1 \rfloor}
\newcommand{\norm}[1]{\mbox{$\left\| #1 \right\|$}}

\newtheorem{theo}{Theorem}
\newtheorem{prop}[theo]{Proposition}
\newtheorem{hypo}[theo]{Hypothesis}
\newtheorem{lemm}[theo]{Lemma}
\newtheorem{conj}[theo]{Conjecture}
\newtheorem{coro}[theo]{Corollary}
\newtheorem{defi}[theo]{Definition}
\newtheorem{rema}[theo]{Remark}
\newtheorem{assu}[theo]{Assumption}
\newtheorem{exam}[theo]{Example}
\newcommand{\theorem}[2][]{\begin{theo}[#1]#2\end{theo}}

\newcommand{\lemma}[1]{\begin{lemm}#1\end{lemm}}

\newcommand{\corollary}[2][]{\begin{coro}[#1]#2\end{coro}}
\newcommand{\definition}[2][]{\begin{defi}[#1]#2\end{defi}}
\newcommand{\rem}[1]{\begin{rema}#1\end{rema}}
\newcommand{\ass}[2][]{\begin{assu}[#1]#2\end{assu}}
\newcommand{\example}[1]{\begin{exam}#1\end{exam}}

\renewcommand{\t}{\tilde}
\newcommand{\h}{\hat}
\newcommand{\ubar}[1]{\underline{#1}}

\newcommand\copyrighttext{%
	\footnotesize 
	\textcopyright 2025 IEEE. Personal use of this material is permitted.
	Permission from IEEE must be obtained for all other uses, in any current or future media, including reprinting/republishing this material for advertising or promotional purposes, creating new collective works, for resale or redistribution to servers or lists, or reuse of any copyrighted component of this work in other works.
}
\newcommand\copyrightnotice{%
	\begin{tikzpicture}[remember picture,overlay]
		\node[anchor=south,yshift=10pt] at (current page.south) {\fbox{\parbox{\dimexpr\textwidth-\fboxsep-\fboxrule\relax}{\copyrighttext}}};
	\end{tikzpicture}%
}

\begin{document}

\title{The Cesàro Value Iteration: Extended Version}

\author{
Jonas Mair$^{a}$,
Lukas Schwenkel$^{a}$,
Matthias A. M\"uller$^{b}$,
Frank Allg\"ower$^{a}$
\thanks{L. Schwenkel thanks the International Max Planck Research School for Intelligent Systems (IMPRS-IS) for supporting him.}
\thanks{\emph{Acknowledgment:} We thank the anonymous reviewer who brought the topic of average reward value iteration \cite{bather1973a}, \cite{hordijk1975modified} to our attention.}
\thanks{$^{a}$ University of Stuttgart, Institute for Systems Theory and Automatic Control, 70550 Stuttgart, Germany; (e-mail: {\it jonas.mair@ist.uni-stuttgart.de, lukas.schwenkel@ist.uni-stuttgart.de, frank.allgower@ist.uni-stuttgart.de})}
\thanks{$^{b}$  Leibniz University Hannover, Institute of Automatic Control, 30167 Hannover, Germany; (e-mail: {\it mueller@irt.uni-hannover.de})}
}

\maketitle
\copyrightnotice
\thispagestyle{empty}

\begin{abstract}	
	In this paper, we consider undiscouted infinite-horizon optimal control for deterministic systems with an uncountable state and input space. We specifically address the case when the classic value iteration does not converge. For such systems, we use the Cesàro mean to define the infinite-horizon optimal control problem and the corresponding infinite-horizon value function. Moreover, for this value function, we introduce the Cesàro value iteration and prove its convergence for the special case of systems with periodic optimal operating behavior. For this instance, we also show that the Cesàro value function recovers the undiscounted infinite-horizon optimal cost, if the latter is well-defined.
\end{abstract}

\section{Introduction}

The Value iteration (VI) is a dynamic programming method and of great importance in the field of optimal control and reinforcement learning (see, e.g., \cite{Bertsekas2005vol1}, \cite{Sutton1998}). However, the classic undiscounted VI does generally not converge to the optimal infinite-horizon cost, or does not converge at all.
This can even happen if the optimal asymptotic average performance is zero and the VI remains bounded (see Example~\ref{ex:motivating_example}). 
In the standard optimal control literature, the convergence problem is often tackled by introducing an exponential discount factor $\ga\in(0,1)$ (e.g. \cite{Bertsekas2005vol1}) in the cost of the underlying optimal control problem (OCP). However, a discount factor $\ga\in(0,1)$ can change the solution of the VI and lead to sub-optimal undiscounted performance. 
Another concept in optimal control is the notion of directly minimizing an asymptotically averaged cost, e.g. \cite{Colonius1989}, \cite{lehrer1992}, \cite{Gaitsgory17}. However, trajectories which are optimal in their asymptotic average can still be arbitrarily bad on any finite interval, since in this case transient costs vanish in the limiting process. A comparison of discounted and asymptotically averaging cost objectives is presented in \cite{dewanto2022examining}.

To overcome the shortcomings of the previous problem formulations, we use the Cesàro mean \cite{Hardy1949} over the stage cost to define an infinite-horizon value function, which we call Cesàro value function. 
The Cesàro mean of a convergent series is consistent with the limit of the partial sums \cite{Hardy1949}. 
For divergent series, the Cesàro mean provides an average value of the partial sums if it exists. 
Furthermore, if the Cesàro mean exists, it is also consistent with the Abelian summation, i.e., the limit $\ga\to1$ of the infinite exponentially discounted sum \cite{Hardy1949}.
We also introduce the Cesàro value iteration (CVI), which allows to recursively calculate the Cesàro value function. The CVI corresponds to VI where the discount factor $\ga$ is modified in each iteration. In particular, in the $N$-th iteration, set $\gamma =\tfrac{N-1}{N}$. 
This VI version was also examined in \cite{bather1973a} and \cite{hordijk1975modified} for Markov decision processes (MDP) in the finite state case and in the context of average reward cost objectives. We show that this choice of gamma arises by defining the cost function via the Cesaro mean, which provides a natural interpretation. In contrast to \cite{bather1973a},\cite{hordijk1975modified}, we consider an uncountable infinite state and input space in a deterministic setup.  

We show that the CVI converges for systems with periodic optimal operating behavior, if a certain dissipativity assumption is satisfied. Furthermore, we show that CVI converges to the optimal infinite horizon cost. A numerical example reveals that the CVI can even have better convergence properties than a VI with constant discount $\ga\in(0,1)$ in the sense that the resulting optimal policy converges faster.

Analyzing convergence of the VI for uncountable infinite state and input spaces is especially relevant in the context of economic model predictive control (EMPC) \cite{Faulwasser2018}, as the value function of an EMPC scheme with horizon $N$ corresponds to the $N$-th iterate of the VI starting at the terminal cost of the EMPC scheme. If the VI does not converge, then the EMPC scheme cannot be expected to achieve optimality with growing prediction horizons as was observed in \cite{Mueller2016}.
One solution is to use linear  discounts \cite{Schwenkel2024}. We show that the linearly discounted cost is exactly the Cesàro cost. Hence, this work provides an intuitive understanding why the scheme of \cite{Schwenkel2024} works and it suggests that the scheme recovers transient optimal performance \cite{Gruene2014} as the prediction horizon tends to infinity (in \cite{Schwenkel2024} only asymptotic average performance guarantees are shown). In \cite{Schwenkel2024a}, the results of \cite{Schwenkel2024} have been extended to a class of non-linear discounts, for which we show in this work that they converge to the Cesàro value function with growing horizon and thus enjoy the same properties. 

\emph{Outline:} In Section \ref{sec:problem}, we propose the Cesàro value function and the CVI, before stating a convergence result thereof in Section \ref{sec:convergence}. In Section \ref{sec:discounts} we introduce a class of cost functions, which contains the Cesàro cost as a special case, and prove that they approximate the Cesàro value function with growing horizons. In Section \ref{sec:numerics} we present a numerical example and in Section \ref{sec:conclusion} a conclusion.

\emph{Notation:} For $a\leq b$, the set of integers in the interval $[a,b]$ is denoted $\SI_{[a,b]}$. We denote the modulo operation by $[k]_p$, i.e., the remainder when dividing $k$ by $p$. For $x\in\R$, the floor operator $\floor{x}$ crops all decimal places. For a continuous function $\al:[0,\infty)\to[0,\infty)$, we say that $\al\in\CK_\infty$ if and only if $\al$ satisfies $\al(0)=0$, $\lim_{t\rightarrow\infty}\al(t)=\infty$ and $\al$ is strictly increasing and for $\de:\N\to[0,\infty)$, we say $\de\in\CL$ if and only if $\de$ is monotonically decreasing and it satisfies $\lim_{t\rightarrow\infty}\de(t)=0$. For a continuous function $h:\SY\to\R$ with a compact domain $\SY\subseteq\R^n$ we abbreviate $h_{\max}\coloneq \max_{y\in\SY}h(y)$ and  $h_{\min}\coloneq\min_{y\in\SY}h(y)$. The cardinality of a set $\SX$ is denoted by $\#\SX$


\section{Cesàro value iteration} \label{sec:problem} 
Consider the nonlinear discrete-time system
\begin{gather}\label{eq:sys}
	x(k+1)=f(x(k),u(k)),
\end{gather}
subject to state and input constraints $x(k)\in\SX\subset\R^n$ and $u(k)\in\SU\subset\R^m$.	A state trajectory of \eqref{eq:sys} resulting from an initial condition $x_0\in\SX$ and a specific input sequence $u\in\SU^N$ of length $N\in\N$ is denoted $x_u(k,x_0)$ and defined by $x_u(0,x_0)=x_0$ and $x_u(k+1,x_0)=f(x_u(k,x_0),u_k)$ for $k\in\SI_{[0,N-1]}$. The set of all feasible control sequences of length $N$ and starting at $x\in\SX$ is denoted $\SU^N(x)\coloneq\{u\in\SU^N~|~\forall k\in\SI_{[0,N]}:x_u(k,x)\in\SX\}$. Furthermore, we have the stage cost $\ell:\SX\times\SU\rightarrow\R$.

\ass[Continuity and compactness]{ \label{ass:cont_comp}
	The functions $f$ and $\ell$ are continuous, the constraints $\SX\times\SU$ are compact and $\SU^N(x)$ is nonempty for all $x\in\SX$ and $N\in\N$.
}
For the stage cost $\ell$, we define the optimal asymptotic average performance
\begin{gather*}
	\ell^\star\coloneq\inf_{u\in\SU^\infty(x)}\liminf_{N\to\infty}\frac{1}{N}\sum_{k=0}^{N-1}\ell(x_u(k,x),u(k))
\end{gather*}
and the optimal infinite-horizon cost as 
\begin{equation}\label{eq:finite_vs_infinite}
	\bar V_\infty(x)\coloneq\inf_{u\in\SU^\infty(x)}\limsup_{N\to\infty}\sum_{k=0}^{N-1}\left(\ell(x_u(k,x),u(k))-\ell^\star\right).
\end{equation}
Since the infinite-horizon problem \eqref{eq:finite_vs_infinite} is in general not tractable, often the classic finite-horizon OCP
\begin{gather*}
	\bar V^1_N(x) \coloneq \inf_{u\in\SU^N(x)} \sum_{k=0}^{N-1}\left(\ell(x_u(k,x),u(k))-\ell^\star\right)
\end{gather*}
and its limit $N\to\infty$ are considered instead. However, the limit $N\to\infty$ of $\bar V^1_N$ may not exist (cf. Example~\ref{ex:motivating_example}), and even if it exists, it may not be equal to $V_\infty(x)$ (cf. Example~\ref{ex:infinite_horizon}).
\example{ \label{ex:motivating_example}
	Consider the system illustrated in Fig. \ref{Fig:Example}, which depicts a system with optimal asymptotic average performance $\ell^\star=0$. One can quickly calculate that $\bar V^1_{N}(x_3)=-1.9$ if $N$ is even and $\bar V^1_{N}(x_3)=0$ if $N$ is odd. Hence, the limit of $\bar V_N^1(x_3)$ for $N\to\infty$ does not exist. Further, for even horizons $N$, it is optimal to start with $u(0)=u_{31}$ and for odd horizons $u(0)=u_{32}$ is optimal. Thus, neither $\bar V^1_N(x)$, nor the corresponding optimal policy converge for $N\to\infty$. 
}

Even if the limit $N\to\infty$ of $\bar V^1_N(x)$ does exist, it might not correspond to the optimal infinite-horizon cost \eqref{eq:finite_vs_infinite}.
\example{  \label{ex:infinite_horizon}
	Consider the example in Fig.~\ref{Fig:Example2}. The optimal behavior over infinite horizons in the sense of \eqref{eq:finite_vs_infinite} is to converge to either $x_1$ or $x_2$ and then stay there. Consequently, when starting at $x_0$, it is better to go to $x_2$ than to $x_1$. However, for $N\geq2$, any finite-horizon optimal trajectory in the sense of $\bar V^1_N(x)$ prefers going to $x_1$ as then, in the last step it can go to $x_3$ with a cost of $-3$. This is impossible in the infinite horizon case as there is no last step. Hence, \eqref{eq:finite_vs_infinite} delivers an optimal cost of $\bar V_\infty(x_0)=1$, while $\bar V^1_N(x_0)=-1$ for $N\geq2$ and therefore also for $N\to\infty$.
}

Motivated by Examples~\ref{ex:motivating_example} and \ref{ex:infinite_horizon}, as a first contribution, we propose to use a Cesàro mean over the shifted stage cost $\bar\ell\coloneq\ell-\ell^\star$ to define an infinite-horizon value function.
\begin{figure}\center
	\begin{tikzpicture}[xscale=1,yscale=1]
		\node[node] (n3) at (0,0)  {$x_3$};
		\node[node] (n1) at (-3.5*\dh,1.75*\dh)  {$x_1$};
		\node[node] (n2) at (3.5*\dh,1.75*\dh)  {$x_2$};
		
		\draw[->] (n2) to [out= 160, in= 20] node[midway,above] {$\ell(x_2,u_{21})=2$} (n1);
		\draw[->] (n1) to [out=-20, in=-160] node[midway,above] {$\ell(x_1,u_{12})=-2$} (n2);

		\draw[->] (n3) to [out= 170, in= -50] node[midway,below] {$\ell(x_3,u_{31})=0.1~~~~~~~~~~~~~~~~$} (n1);
		\draw[->] (n3) to [out= 10, in= -130] node[midway,right] {$~~~~~\ell(x_3,u_{32})=0$} (n2);
		
	\end{tikzpicture}
	\caption{
		Illustration of the states $x_{i}$ (nodes) and feasible transitions (edges) with corresponding inputs $u_{ij}$ and costs $\ell(x_i,u_{ij})$ for a system which is optimally operated at a periodic orbit with average cost $\ell^\star=0$.
	}
	\label{Fig:Example}
\end{figure}
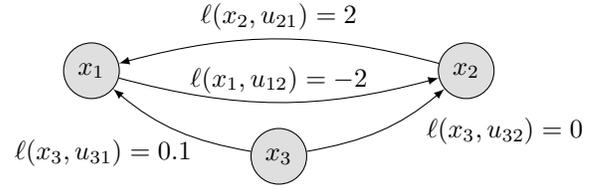
\begin{figure}\center
	\begin{tikzpicture}[xscale=1,yscale=1]
		\node[node] (n1) at (0,0)  {$x_1$};
		\node[node] (n0) at (-2.5*\dh,-1.75*\dh)  {$x_0$};
		\node[node] (n3) at (2.5*\dh,-1.75*\dh)  {$x_3$};
		\node[node] (n2) at (0,-3.5*\dh)  {$x_2$};
		
		\draw[->] (n1) to [out=125, in=55, looseness=7] node[midway,above] {$\ell(x_1,u_{11})=0$} (n1);
		\draw[->] (n2) to [out= 305, in= 235, looseness=7] node[midway,below] {$\ell(x_2,u_{22})=0$} (n2);

		\draw[->] (n0) to [out= 65, in= 180] node[left=0.1cm] {$\ell(x_3,u_{01})=2$} (n1);
		\draw[->] (n1) to [out= 0, in= 115] node[right=0.1cm] {$\ell(x_3,u_{13})=-3$} (n3);
		\draw[->] (n0) to [out= 295, in= 180] node[left=0.1cm] {$\ell(x_3,u_{02})=1$} (n2);	
		\draw[->] (n2) to [out= 0, in= 245] node[right=0.1cm] {$\ell(x_3,u_{23})=-1$} (n3);
		\draw[->] (n3) to [out= 180, in= 0] node[above] {$\ell(x_3,u_{30})=5$} (n0);
	\end{tikzpicture}
	\caption{
		Illustration of the states $x_{i}$ (nodes) and feasible transitions (edges) with corresponding inputs $u_{ij}$ and costs $\ell(x_i,u_{ij})$ for a system which is optimally operated at a periodic orbit with average cost $\ell^\star=0$.
	}
	\label{Fig:Example2}
\end{figure}
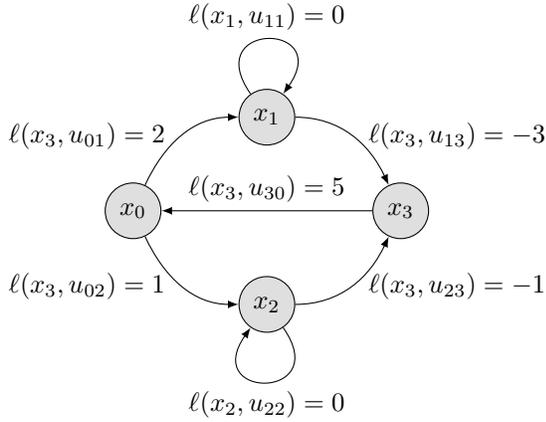
%
\begin{table}\centering 
	\caption{Values of the value functions for the example in Fig.\ref{Fig:Example2}.}
	\begin{tabular}{c||c|c|c}
		& $\lim\limits_{N\to\infty}\bar V^1_N(x)$ & $\bar V^\mathrm{ces}_\infty(x)$ & $\bar V_\infty(x)$\\ \hline\hline
		$x_0$ & $-1$ & $1$ & $1$\\ \hline
		$x_1$ & $-3$ & $0$ & $0$\\ \hline
		$x_2$ & $-1$ & $0$ & $0$\\ \hline
		$x_3$ & $4$ & $6$ & $6$
	\end{tabular}
	\label{tab:vf}
\end{table}
\definition[Cesàro Value Function]{
	The averaged sum
	\begin{gather*}
		\bar J^\mathrm{ces}_N(x,u)\coloneq\frac{1}{N}\sum_{n=0}^{N-1}\sum_{k=0}^{n}\bar\ell(x_u(k,x),u(k))
	\end{gather*}
	over the shifted stage cost $\bar\ell$ is called the Cesàro cost and the corresponding OCP
	\begin{gather*}
		\bar V^\mathrm{ces}_N(x)\coloneq\inf_{u\in\SU^N(x)} \bar J^\mathrm{ces}_N(x)
	\end{gather*}
	the finite-horizon Cesàro value function. If the limit exists, we also define the infinite-horizon Cesàro value function by the Cesàro mean $\bar V^\mathrm{ces}_\infty(x)\coloneq\lim_{N\rightarrow\infty}\bar V^\mathrm{ces}_N(x)$. 
}
%
\rem{
	The Cesàro cost $\bar J^\mathrm{ces}_N$ is equivalent to the linearly discounted cost functional from \cite{Schwenkel2024}, which can be seen by the algebraic reformulation
	\begin{gather} \label{eq:ces_lin}
		\frac{1}{N}\sum_{n=0}^{N-1}\sum_{k=0}^{n} \bar\ell_k 
		= \frac{1}{N}\sum_{k=0}^{N-1} (N-k) \bar\ell_k
		= \!\sum_{k=0}^{N-1} (1-\frac{k}{N}) \bar\ell_k,
	\end{gather}
	where the abbreviation $\bar\ell_k\coloneq\bar\ell(x_u(k,x),u(k))$ was used.
}

Note that due to Assumption~\ref{ass:cont_comp}, there exists a minimizing sequence $\bar u^\mathrm{ces}_{N,x}$ that attains the infimum, i.e., $\bar V^\mathrm{ces}_N(x)=\bar J^\mathrm{ces}_N(x,\bar u^\mathrm{ces}_{N,x})$.

The following two examples demonstrate that the Cesàro mean recovers the optimal infinite-horizon cost $\bar V_\infty(x)$ in \eqref{eq:finite_vs_infinite}.
\example{
	For the system in Fig. \ref{Fig:Example}, in contrast to the classic value function $\bar V^1_N$, the Cesàro value function converges for $N\to\infty$ to $\bar V^\mathrm{ces}_\infty(x_1)=-1$, $\bar V^\mathrm{ces}_\infty(x_2)=1$ and $\bar V^\mathrm{ces}_\infty(x_3)=-0.9$. Moreover, the Cesàro value function provides the infinite-horizon optimal policy, i.e., at $x_3$ it is optimal to apply $u_{31}$.}

\example{ \label{ex:infinite_horizon_2}
	For the system in Fig.~\ref{Fig:Example2}, minimizing the Cesàro mean over the transition costs yields $\bar V^\mathrm{ces}_N(x_0)=1-\tfrac{1}{N}$ for $N\geq2$ and hence $\bar V^\mathrm{ces}_\infty(x_0)=1$, which corresponds to applying the infinite-horizon optimal input $u_{02}$ at $x_0$ and then staying at $x_2$. The reason is that the cost of the last step becomes irrelevant for large $N$ due to the discount. An overview of the resulting value functions is shown Table~\ref{tab:vf}. As a conclusion, the Cesàro mean $\bar V^\mathrm{ces}_\infty$ calculates the infinite-horizon costs \eqref{eq:finite_vs_infinite}, whereas the undiscounted value function $\lim_{N\to\infty} \bar V^1_N$ does not.
}

The fact that $\bar V_\infty(x)=\bar V^\mathrm{ces}_\infty(x)$ holds in Example~\ref{ex:infinite_horizon_2} is no coincidence. 
If $\bar V_\infty(x)$ in \eqref{eq:finite_vs_infinite} has a well-defined limit for all $x\in\SX$ in the sense that it satisfies 
\begin{align}\label{eq:infinite_horizon_cost}
	\bar V_\infty(x)=\inf_{u\in\SU^\infty(x)}\liminf_{N\to\infty}\sum_{k=0}^{N-1}\ell(x_u(k,x),u(k)),
\end{align}
then, under certain assumptions, it can be proven that $\bar V_\infty(x)=\bar V^\mathrm{ces}_\infty(x)$ for all $x\in\SX$ (cf. Theorem~\ref{lem:difference_rotated_value_fcns}).
%

Our main conceptual contribution is a recursive procedure to compute the Cesàro value function: the Cesàro value iteration.
\theorem[Cesàro value iteration]{\label{thm:value_iteration}
	Let Assumption \ref{ass:cont_comp} hold. For all $x\in\SX$ and all $N\in\N$, the finite-horizon Cesàro value function $\bar V^\mathrm{ces}_N(x)$ satisfies
	\begin{gather}\label{eq:value_iteration}
		\bar V^\mathrm{ces}_N(x)=\min_{u\in\SU^1(x)} \bar\ell(x,u) + \frac{N-1}{N} \bar V^\mathrm{ces}_{N-1}(f(x,u)).
	\end{gather}
}
\begin{proof}
	For a general $u\in\SU^N(x)$, we have
	\begin{flalign}
		\begin{split}
			\bar J^\mathrm{ces}_N&(x,u)= \bar\ell(x,u(0)) + \sum_{k=1}^{N-1}\frac{N-k}{N}\ell(x_u(k,x),u(k)) \\
			=&\bar\ell(x,u(0)) + \frac{N-1}{N}\sum_{k=1}^{N-1}\frac{N-k}{N-1}\ell(x_u(k,x),u(k)) \\
			=&\bar\ell(x,u(0))+\frac{N-1}{N}\bar J^\mathrm{ces}_{N-1}(f(x,u(0)),u_{[1,N-1]}),  
		\end{split}\hspace{-0.3cm}&\label{eq:rec_cost}
	\end{flalign}
	where $u_{[1,N-1]}(k-1)=u(k)$ for $k\in\SI_{[1,N-1]}$. We prove equality in \eqref{eq:value_iteration} by showing $\leq$ and $\geq$ starting with $\leq$. For any $u(0)\in\SU^1(x)$, we choose the input sequence $u_{[1,N-1]}=\bar u^\mathrm{ces}_{N-1,f(x,u(0))}$ in \eqref{eq:rec_cost}. Taking the minimum over $u(0)$ and exploiting $\bar V^\mathrm{ces}_N(x)\leq \bar J^\mathrm{ces}_N(x,u)$ (by optimality), this already shows \eqref{eq:value_iteration} with $\leq$. For the case $\geq$, a different input sequence $u=\bar u^\mathrm{ces}_{N,x}$ is considered. Optimality yields $\bar V^\mathrm{ces}_{N-1}(f(x,u(0))\leq \bar J^\mathrm{ces}_{N-1}(f(x,u(0)),u_{[1,N-1]})$ and with \eqref{eq:rec_cost} and taking the minimum over $u(0)$ this leads to \eqref{eq:value_iteration} with $\geq$, which concludes the proof.
\end{proof}
\rem{
	The iteration scheme in Theorem \ref{thm:value_iteration} is the deterministic version of the scheme presented in \cite{bather1973a}. 
}
\rem{ \label{rem:ellbarell}
	The CVI can also be defined by substituting $\bar\ell$ by $\ell$ and accordingly defining $V^\mathrm{ces}_N$. Note that the minimizer $u^\mathrm{ces}_{N,x}$ satisfies $\bar u^\mathrm{ces}_{N,x}= u^\mathrm{ces}_{N,x}$, since the cost is shifted by a constant. For convergence guarantees of $V^\mathrm{ces}_N$, see Remark~\ref{rem:convergence}.
}

\corollary[\it Bellman Equation]{
	If $\bar V^\mathrm{ces}_\infty(x)$ exists for all $x\in\SX$, then it satisfies the Bellman equation
	\begin{gather*}
		\bar V^\mathrm{ces}_\infty(x)=\min_{u\in\SU^1(x)} \bar\ell(x,u) + \bar V^\mathrm{ces}_\infty(f(x,u)).
	\end{gather*}
}



\section{Convergence of the Cesàro Value Iteration} \label{sec:convergence}

In this section, we present our main theoretical contribution: a convergence guarantee for the CVI under certain assumptions. In particular, for the large class of systems with optimal periodic operation (which includes optimal steady-state operation as a special case) we can prove its convergence. Let us first formally define (optimal, minimal) periodic orbits. 
\definition[{\it Optimal Periodic Orbit,} {\cite[Def. 2]{Schwenkel2024}}]{\label{def:optimal_periodic_orbit}
	A $p$-tuple $\varPi=((x_0,u_0),\ldots,(x_{p-1},u_{p-1}))\in (\SX\times\SU)^p$, $p\in\N$ is called a feasible $p$-periodic orbit, if its projection $\varPi_\SX\coloneq(x_0,\ldots,x_{p-1})$ onto $\SX^p$ satisfies for all $k\in\SI_{[0,p-1]}$
	\begin{gather*}
		\varPi_\SX([k + 1]_p) = f (\varPi(k)).
	\end{gather*}
	A $p$-periodic orbit $\varPi$ is called minimal, if $\varPi_\SX(k)=\varPi_\SX(j) \Rightarrow k=j$ for all $k,j\in\SI_{[0,p-1]}$. The distance of a pair $(x,u)\in\SX\times\SU$ to the orbit $\varPi$ is defined as $\norm{(x,u)}_\varPi\coloneq\inf_{k\in\SI_{[0,p-1]}} \norm{(x, u)-\varPi(k)}$. The set of all feasible $p$-periodic orbits is denoted by $S_\varPi^p$. The average cost at $\varPi\in S_\varPi^p$ is defined as $\ell^p(\varPi) \coloneq \frac{1}{p}\sum_{k=0}^{p-1}\ell(\varPi(k))$. If a feasible $p^\star$-periodic orbit $\varPi^\star$	satisfies
	\begin{gather} \label{eq:opt_orbit}
		\ell^{p^\star}(\varPi^\star)=\inf_{p\in\N,\varPi\in S_\varPi^p} \ell^p(\varPi)
	\end{gather}
	then $\varPi^\star$ is called an optimal periodic orbit and $p^\star$ is called an optimal period length.
}

A minimal periodic orbit corresponds to the definition of a cycle in graph theory. Thus, a minimal orbit cannot be 8-shaped or contain multiple laps. Note that \emph{minimality} refers to the length of an orbit, whereas \emph{optimality} refers to its average cost. Minimality is also part of the following assumptions, known from the EMPC literature \cite{Mueller2016}, \cite{Schwenkel2024}, \cite{Schwenkel2024a}, that we need in order to show convergence of the CVI. 

\ass[Strict dissipativity]{\label{ass:strict_dissipativity}
	There exist a minimal optimal periodic orbit $\varPi^\star$, a continuous storage function $\lb:\SX\rightarrow\R$, $\bar{\lb}\in\R$ and a function $\al_{\t\ell}\in\CK_\infty$, such that $\abs{\lb(x)}\leq\bar{\lb}$ for all $x\in\SX$ and such that the rotated stage cost satisfies
		$\t\ell(x,u)\coloneq\bar\ell(x,u)+\lb(x)-\lb(f(x,u))\geq\al_{\t\ell}(\norm{(x,u)}_{\varPi^\star})$
	for all $x\in\SX$ and $u\in\SU^1(x)$.
}

As noted in \cite{Mueller2015}, strict dissipativity implies optimality of $\varPi^\star$ and that $\ell^\star=\ell^{p^\star}(\varPi^\star)$, i.e., that the optimal asymptotic average performance is equal to the average cost of the optimal periodic orbit $\varPi^\star$. 
Conversely, if the system \eqref{eq:sys} is locally controllable\footnote{cf. Assumption \ref{ass:local_ctrb}} and uniformly suboptimally operated off the periodic orbit $\varPi^\star$, i.e., if operation at the optimal periodic orbit $\varPi^\star$ is strictly better than any other operating behavior, then the system is strictly dissipative w.r.t. the periodic orbit $\varPi^\star$ \cite{Mueller2015}.
We also require two controllability conditions stated in \cite[Ass. 10 and 11]{Mueller2016}. These assumptions imply that $\varPi^\star$ can be reached in finite time from all initial conditions $x_0\in\SX$.
%
\ass[Local controllability at $\varPi^\star$]{\label{ass:local_ctrb}
	There exist $\kappa>0$, $M'\in\N$ and $\rho\in\CK_\infty$ such that for all $z\in\varPi_\SX^\star$ and all $x,y\in\SX$ with $\norm{x-z}\leq\kappa$ and $\norm{y-z}\leq\kappa$ there exists a control input sequence $u\in\SU^{M'}(x)$ that satisfies $x_u(M',x)=y$ and
		$\norm{(x_u(k,x),u(k))}_{\varPi^\star}\leq\rho(\max\{\norm{x}_{\varPi_\SX^\star},\norm{y}_{\varPi_\SX^\star}\})$
	for all $k\in\SI_{[0,M'-1]}$.
}
\ass[Finite-time reachability of $\varPi^\star$]{ \label{ass:finite_reachability}
	For $\kappa>0$ from Assumption \ref{ass:local_ctrb} there exists $M''\in\SI_{\geq1}$ such that for all $x\in\SX$ there exist $K\in\SI_{[0,M'']}$ and $u\in\SU^K(x)$ satisfying $\norm{x_u(K,x)}_{\varPi_\SX^\star}\leq\kappa$.
}

Technically, Assumption \ref{ass:finite_reachability} only implies that a neighborhood of $\varPi^\star$ can be reached. Assumption \ref{ass:local_ctrb} then guarantees that $\varPi^\star$ can be reached exactly from within this neighborhood.
Under these assumptions we can guarantee convergence of the CVI.

\theorem[Convergence of CVI]{\label{thm:cesaro_convergence}
	Let Assumptions \ref{ass:cont_comp}, \ref{ass:strict_dissipativity}, \ref{ass:local_ctrb} and \ref{ass:finite_reachability} hold and let $x\in\SX$.
	Then, $\lim_{N\rightarrow\infty} \bar V^\mathrm{ces}_N(x)$ exists and is finite.
}

We omit the proof of Theorem \ref{thm:cesaro_convergence} here since we will later provide a more general result in Theorem~\ref{thm:convergence} and Theorem~\ref{thm:cesaro_convergence} is a special case thereof. 

\rem{\label{rem:convergence}
	When we do not shift the cost by $\ell^\star$ (cf. Remark~\ref{rem:ellbarell}) and $\ell^\star\neq 0$, then Theorem~13 guarantees convergence of the differences: for all $x\in\SX$, $V_{N}^\mathrm{ces}(x)-V_{N-1}^\mathrm{ces}(x) = \bar V_{N}^\mathrm{ces}(x)-\bar V_{N-1}^\mathrm{ces}(x) + \sum_{k=0}^{N-1} \tfrac{N-k}{N} \ell^\star - \sum_{k=0}^{N-2} \tfrac{N-1-k}{N-1}\ell^\star = \bar V_{N}^\mathrm{ces}(x)-\bar V_{N-1}^\mathrm{ces}(x) +\tfrac{\ell^\star}{2} \to \tfrac{\ell^\star}{2}$ as $N\to\infty$, since 
	$\sum_{k=0}^{N-1}\tfrac{N-k}{N}=\tfrac{N+1}{2}$.
}

For MDPs in the finite state setting, the CVI converges under mild assumptions \cite{bather1973a}. Since we consider an uncountable infinite state space, we need stronger assumptions in Theorem \ref{thm:cesaro_convergence} to prove convergence of the CVI. However, in the deterministic setting with finite states, systems are always optimally operated at a periodic orbit. Hence, as shown in Section~\ref{sec:finite_states}, in this case our assumptions reduce to similar ones as in \cite{bather1973a} with the additional assumption that a unique minimal optimal periodic orbit exists.


\section{Discounted Cost Functionals} \label{sec:discounts}

In this section, we introduce a more general class of cost functions and show that their limit $N\to\infty$ is equal to the Cesàro value function $\bar V^\mathrm{ces}_\infty$. Recall that by \eqref{eq:ces_lin}, the Cesàro cost $\bar J^\mathrm{ces}_N$ is equivalent to the linearly discounted cost functional from \cite{Schwenkel2024}. 
The equivalence of a linearly discounted cost function and the Cesàro cost gives an intuitive explanation why the EMPC scheme in \cite{Schwenkel2024} works. It minimizes the Cesàro cost online and for growing prediction horizons approximates the infinite-horizon Césaro value function. Therefore, it achieves transient optimal performance for $N\to\infty$. The linear discounts have been extended to a more general class of discount functions $\be:[0,1]\to[0,1]$ in \cite{Schwenkel2024a}. 

\ass[{Discount function}, {\cite[Ass. 5]{Schwenkel2024a}}]{\label{ass:discount_fcn}
	The discount function $\be:[0,1]\rightarrow[0,1]$ is piece-wise continuously differentiable, Lipschitz-continuous with Lipschitz constant $L_\be$, non-increasing, and satisfies $\be(0)=1$, $\be(1)=0$ and $\limsup_{\xi\rightarrow1}\be'(\xi)<0$.
}

To establish the same intuitive understanding for the EMPC scheme in ~\cite{Schwenkel2024a}, we show our results for such a discounted cost functional
\begin{gather}\label{eq:discounted_cost}
	\bar J^\be_N(x,u) \coloneq \sum_{k=0}^{N-1}\be(\tfrac{k}{N})\bar\ell(x_u(k,x),u(k))
\end{gather}
and the discounted finite-horizon OCP
\begin{gather}\label{eq:discounted_value_fcn}
	\bar V^\be_N(x) \coloneq \inf_{u\in\SU^N(x)} \bar J^\be_N(x,u).
\end{gather}

\rem{
	The recursive formula in \eqref{eq:value_iteration} does typically not hold for such general discount functions $\be$. This is because the proof of Theorem \ref{thm:value_iteration} exploits that $\be(\tfrac{k}{N})\be(\tfrac{k-1}{N-1})^{-1}$ is independent of $k$, which is not true for general $\be$. For the linear discount function $\be(\xi)=1-\xi$, however, we have $\be(\tfrac{k}{N})\be(\tfrac{k-1}{N-1})^{-1}=\tfrac{N-1}{N}$.
}

To prove that the limit $N\to\infty$ of the discounted value function $\bar V^\be_N$ exists, we rely on the well-known notion of rotated costs \cite{Faulwasser2018}. 
If Assumption \ref{ass:strict_dissipativity} is satisfied, any storage function $\lb$ satisfying the assumption can be used to define the rotated stage cost $\t\ell$. The rotated cost functional $\t J^\be_N$ and the rotated value function $\t V^\be_N$ are then obtained by replacing $\bar\ell$ with $\t\ell$ in \eqref{eq:discounted_cost} and \eqref{eq:discounted_value_fcn}. We emphasize that to apply the CVI one does not need to explicitly find a storage function $\lb$, since the rotated quantities are only required for the theoretical proofs and not for the eventual application of the algorithm.
The undiscounted case is denoted by $\be \equiv 1$.
Since $\t\ell$ and $\bar\ell$ are continuous due to Assumptions~\ref{ass:cont_comp} and \ref{ass:strict_dissipativity}, minimizing sequences $\bar u^\be_{N,x}$, $\t u^\be_{N,x}$ and $\t u^1_{N,x}$ exist.
As an intermediate result, we first prove the existence of the limit $N\to\infty$ for the rotated discounted value function $\t V^\be_N$.

\lemma{ \label{lem:convergence_rotated_value_fcn}
	Let Assumptions \ref{ass:cont_comp}, \ref{ass:strict_dissipativity}, \ref{ass:local_ctrb}, \ref{ass:finite_reachability} and \ref{ass:discount_fcn} be satisfied. For each $x\in\SX$, the limit $\lim_{N\rightarrow\infty}\t V^\be_N(x)\eqcolon\t V^\be_\infty(x)$ exists and is finite.
}
\begin{proof} 
	We use Lemma \ref{lem:uniform_bound} (upper bound on $\t V^\be_N(x)$) in the Appendix and Lemma \ref{lem:monotonicity_rot_value_fcn} (monotonicity of $\t V^\be_N(x)$) in the Appendix to infer convergence to a finite limit.
\end{proof}

In contrast to $\t\ell$, the shifted stage cost $\bar\ell$ might be negative. Hence, we cannot use the same arguments to show convergence of $\bar V^\be_N$. However, we can relate the shifted and the rotated value function.

\lemma{ \label{lem:convergence_value_fcn}
	Let Assumptions  \ref{ass:cont_comp}, \ref{ass:strict_dissipativity}, \ref{ass:local_ctrb}, \ref{ass:finite_reachability}, \ref{ass:discount_fcn} hold and let $x\in\SX$. Then there exists $\de\in\CL$ such that
	\begin{gather} \label{eq:difference_value_fcns}
		\abs{\t V^\be_N(x) - \bar V^\be_N(x) - \lb(x)+\lb_{\varPi^\star}}\leq\de(N),
	\end{gather}
	for all $N\in\N$, where $\lb_{\varPi^\star}\coloneq\tfrac{1}{p^\star}\sum_{k=0}^{p^\star-1}\lb(\varPi^\star(k))$.
}
\begin{proof}
	By optimality, we have $\t V^\be_N(x)\leq \t J^\be_N(x,\bar u^\be_{N,x})$. Hence, with Lemma \ref{lem:rotated_cost_fcn} from the Appendix, we infer
	\begin{align*}
		\t V^\be_N(x)&-  \bar V_N^\be(x)
		\leq \t J^\be_N(x,\bar u^\be_{N,x})- \bar J^\be_N(x,\bar u^\be_{N,x}) \\
		&= \lb(x)-\sum_{k=1}^{N}(\be(\tfrac{k-1}{N})-\be(\tfrac{k}{N}))\lb(x_{\bar u^\be_{N,x}}(k,x)).
	\end{align*}
	By Assumptions~\ref{ass:strict_dissipativity} and  \ref{ass:discount_fcn}, we infer that $-\lb(x_{u^\be_{N,x}}(N,x))\be(\tfrac{N-1}{N})\leq \bar\lb\be(\tfrac{N-1}{N})\eqcolon \de_1(N)\in\CL$. Hence, under Assumptions \ref{ass:cont_comp}, \ref{ass:strict_dissipativity}, \ref{ass:local_ctrb}, \ref{ass:finite_reachability} and \ref{ass:discount_fcn}, we can apply \cite[Lemma~4]{Schwenkel2024a}, which exploits turnpike arguments, to guarantee the existence of some $\de_2\in\CL$ such that
	$-\sum_{k=1}^{N}(\be(\tfrac{k-1}{N})-\be(\tfrac{k}{N}))\lb(x_{u^\be_{N,x}}(k,x)) \leq \de(N) -\lb_{\varPi^\star},$
	with $\de(N)\coloneq\de_1(N)+\de_2(N)$, and hence
	\begin{gather*}
		\t V^\be_N(x)-\bar V_N^\be(x)-\lb(x)+\lb_{\varPi^\star}\leq \de(N).
	\end{gather*}
	By starting with $\bar V^\be_N(x)\leq \bar J^\be_N(x,\t u^\be_{N,x})$ by optimality, we can apply similar arguments as above to also obtain
	\begin{gather*}
		-(\t V^\be_N(x)-\bar V_N^\be(x)-\lb(x)+\lb_{\varPi^\star})\leq \de(N),
	\end{gather*}
	which proves \eqref{eq:difference_value_fcns}. 
\end{proof}

Now, we can show convergence of $\bar V^\be_N$.

\theorem[Infinite-horizon discounted value function]{\label{thm:convergence}
	Let Assumptions  \ref{ass:cont_comp}, \ref{ass:strict_dissipativity}, \ref{ass:local_ctrb}, \ref{ass:finite_reachability}, \ref{ass:discount_fcn} hold and let $x\in\SX$.
	Then $\bar V^\be_\infty(x)\coloneq\lim_{N\rightarrow\infty} \bar V^\be_N(x)$ exists and is finite.
}
\begin{proof}
	The existence of $\lim_{N\rightarrow\infty} \bar V^\be_N(x)<\infty$, is a simple consequence of Lemma \ref{lem:convergence_rotated_value_fcn} and \ref{lem:convergence_value_fcn}.
\end{proof}

Not only does $\bar V^\be_N$ converge for $N\to\infty$, but it converges to $\bar V^\mathrm{ces}_\infty$ for all $\be$ that satisfy Assumption \ref{ass:discount_fcn}, as we show next.

\theorem[Value of convergence]{ \label{lem:difference_rotated_value_fcns}
	Let Assumptions~\ref{ass:cont_comp}, \ref{ass:strict_dissipativity}, \ref{ass:local_ctrb}, \ref{ass:finite_reachability} and \ref{ass:discount_fcn} be satisfied. Then, for all $x\in\SX$, the relation $\bar V^\be_\infty(x) = \bar V^\mathrm{ces}_\infty(x)$ holds. If additionally \eqref{eq:infinite_horizon_cost} is satisfied, then also $\bar V_\infty(x)=\bar V^\mathrm{ces}_\infty(x)$ holds for all $x\in\SX$.
}
\begin{proof} 
	Since the proof of the second part, i.e., that $\bar V_\infty(x)=\bar V^\mathrm{ces}_\infty(x)$, requires additional Lemmas, we moved that part to Subsection~\ref{subsec:proof}. Here, we only prove the first part, i.e.,  $\bar V^\be_\infty(x)=\bar V^\mathrm{ces}_\infty(x)$.
	
	First, by optimality, and since $\be(\tfrac{k}{N})\in[0,1]$ for all $N\in\N$ and $\t\ell\geq0$, we have $\t V^\be_N(x)\leq\t J^\be_N(x,\t u^1_{N,x})\leq \t J^1(x,\t u^1_{N,x})=\t V^1_N(x)$, which implies $\liminf_{N\to\infty}\t V^1_N(x)\geq\t V^\be_\infty(x)$.
	Secondly, note that $\sum_{k=0}^{K-1}k\leq K^2$ for $K\in\N$ and hence, by exploiting nonnegativity of $\be$ and $\t\ell$ and Lipschitz continuity of $\be$ (Ass. \ref{ass:discount_fcn}), we obtain for all $N\in\N$
	\begin{align*}
		\t V^\be_N(x)
		&\geq \sum_{k=0}^{\floor{\sqrt[4]{N}}-1} \be(\tfrac{k}{N})\t\ell(x_{u^\be_{N,x}}(x,k),u^\be_{N,x}(k)) \\
		&\geq \sum_{k=0}^{\floor{\sqrt[4]{N}}-1} \left(1-L_\be\tfrac{k}{N}\right)\t\ell(x_{u^\be_{N,x}}(x,k),u^\be_{N,x}(k)) \\
		&\geq \t V^1_{\floor{\sqrt[4]{N}}}(x) - L_\be\tfrac{1}{\sqrt{N}}\t\ell_{\max},
	\end{align*}
	which implies that $\limsup_{N\to\infty}\t V^1_N(x)\leq \t V^\be_\infty(x)$.
	Thus, $\t V^\be_\infty(x)=\lim_{N\to\infty}\t V^1_N(x)$ holds, which proves that, for all $x\in\SX$, $\t V^\be_\infty(x)$ takes the same value for all $\be$ that satisfy Assumption \ref{ass:discount_fcn}.
	By combining this result with Lemma \ref{lem:convergence_value_fcn} and \eqref{eq:ces_lin}, we infer that, for all $x\in\SX$, $\bar V^\be_\infty(x)=\bar V^\mathrm{ces}_\infty(x)$ for all $\be$ that satisfy Assumption \ref{ass:discount_fcn}, since the linear discount function $\be(\xi)=1-\xi$ satisfies Assumption~\ref{ass:discount_fcn}.
\end{proof}


\subsection{Proof of the second part of Theorem~\ref{lem:difference_rotated_value_fcns}} \label{subsec:proof}

%
In this section, the second part of Theorem~\ref{lem:difference_rotated_value_fcns} is proven. In order to do so, first we define $\t V_\infty(x)$ analogously to \eqref{eq:finite_vs_infinite} by replacing $\ell-\ell^\star$ with $\t\ell$. Further, for any $u\in\SU^\infty(x)$, we use the abbreviation $\bar J^1_\infty(x,u)\coloneq\lim_{N\to\infty}\bar J^1_N(x,u)$ if the limit exists. This notation is used for the rotated quantities as well. 

Throughout this section it will be assumed that the minimum in \eqref{eq:finite_vs_infinite} and a corresponding minimizing input $u^1_{\infty,x}\in\SU^\infty(x)$ exist, i.e., $\bar V_\infty(x)=\limsup_{N\to\infty}\bar J^1_N(x,u^1_{\infty,x})$. Similarly, we assume the existence of a minimizing input $\t u^1_{\infty,x}\in\SU^\infty(x)$ for the rotated infinite-horizon cost $\t V_\infty$. We note that these assumptions are only made to avoid technicalities that distract from the main ideas of the proof. One could also work with inputs that only approximate the infinite-horizon cost up to some arbitrary error $\eps>0$ and adapt the arguments in the proofs accordingly. 

We begin the proof by first establishing convergence of the finite-horizon value function $\t V^1_N$ as $N\to\infty$ and showing that the corresponding optimal state at time $N$ converges to the optimal orbit as $N\to\infty$. 
\lemma{ \label{lem:convergence_to_orbit}
	Let Assumptions~\ref{ass:cont_comp}, \ref{ass:strict_dissipativity}, \ref{ass:local_ctrb} and \ref{ass:finite_reachability} be satisfied. Then, for all $x\in\SX$ the limit $\t V^1_\infty(x)\coloneq\lim_{N\to\infty}\t V^1_N(x)$ exists and is finite. Moreover, for all $x\in\SX$ there exists $\de\in\CL$ such that $\|x_{\t u^1_{N,x}}(N)\|_{\varPi^\star_\SX}\leq\de(N)$.
}
\begin{proof}
	To prove the existence of $\lim_{N\to\infty}\t V^1_N(x)$, it is sufficient to show that for all $x\in\SX$ the rotated value function $\t V^1_N$ is upper bounded and that $\t V^1_N$ is monotonically increasing in $N$, i.e., $\t V^1_N(x)\geq\t V^1_{N-1}(x)$. For the upper bound, consider $N\geq M'+M''\eqcolon M$ with $M'$ and $M''$ from Assumptions~\ref{ass:local_ctrb} and \ref{ass:finite_reachability}, respectively. Then, for all $x\in\SX$ we can define a feasible candidate input $u_x\in\SU^N$ by choosing $u_x$ such that $\|x_{u_x}(M,x)\|_{\varPi^\star_\SX}=0$ and $\|u_x(k)\|_{\varPi^\star_\SU}=0$ for $k\in\SI_{[M,N-1]}$. This implies for all $x\in\SX$ and $N\in\N$ that $\t V^1_N(x)\leq\t J^1_N(x,u_x)=\sum_{k=0}^{M-1}\t\ell(x_{u_x}(k,x),u_x(k))\leq M\t\ell_\mathrm{max}$.
	To show the monotonic increase, pick the optimal input $\t u^1_{N,x}\in\SU^N(x)$ to obtain 
	\begin{align}\label{eq:monotone_increase}
		\begin{split}
			\t V^1_N(x)-\t V^1_{N-1}(x) \geq \t J^1_N(x,\t u^1_{N,x})-\t J^1_{N-1}(x,\t u^1_{N,x}) \\
			= \t\ell(x_{\t u^1_{N,x}}(N-1),\t u^1_{N,x}(N-1))\geq0.
		\end{split}
	\end{align}
	Thus, for all $x\in\SX$ the limit $\t V^1_\infty(x)\coloneq\lim_{N\to\infty}\t V^1_N(x)$ exists and is finite. 
	Next, we claim that for all $x\in\SX$ there exists $\h\de\in\CL$ such that 
	\begin{equation}\label{eq:claim}
		\h\de(N)\geq\t\ell(x_{\t u^1_{N,x}}(N-1),\t u^1_{N,x}(N-1)).
	\end{equation}
	This is indeed the case, since we can choose $\h\de(N)\coloneq\t V^1_\infty(x)-\t V^1_{N-1}(x)\geq\t V^1_N(x)-\t V^1_{N-1}(x)$ for $N\in\N$. The inequality holds because $\t V^1_N(x)$ is monotonically increasing in $N$ and converges to $\t V^1_\infty(x)$. Moreover, because of these two properties, $\h\de(N)$ is monotonically decreasing and $\lim_{N\to\infty}\h\de(N)=0$, which proves the claim.
	Since $\t\ell(x,u)\geq\ubar\al_{\t\ell}(\|(x,u)\|_{\varPi^\star})$ for $(x,u)\in\SX\times\SU$, inequality \eqref{eq:claim} implies $\|(x_{\t u^1_{N,x}}(N-1),\t u^1_{N,x}(N-1))\|_{\varPi^\star}\leq\ubar\al^{-1}_{\t\ell}(\h\de(N))$. By Lemma~\ref{lem:follow_orbit} in the Appendix we infer $\|x_{\t u^1_{N,x}}(N)\|_{\varPi^\star_\SX}<\al_f(\al^{-1}_{\t\ell}(\h\de(N)))$ and since $\al_f,\ubar\al_{\t\ell}\in\CK_\infty$, we conclude $\de\coloneq\al_f\circ\al^{-1}_{\t\ell}\circ\h\de\in\CL.$
\end{proof}
%

Next, it is shown that taking the limit $N\to\infty$ of the rotated finite-horizon value function $\t V_N$ delivers the same value as the rotated infinite-horizon value function $\t V_\infty$. We emphasize that for the non-rotated value functions this is not necessarily the case.

%
\lemma{ \label{lem:equiv_rotated_finite_infinte_horizon}
	Let Assumptions~\ref{ass:cont_comp}, \ref{ass:strict_dissipativity}, \ref{ass:local_ctrb} and \ref{ass:finite_reachability} be satisfied Then, for all $x\in\SX$ it holds that
	\begin{equation} \label{eq:rotated_infinite_horizon_cost}
		\t V_\infty(x)=\inf_{u\in\SU^\infty(x)}\liminf_{N\to\infty}\t J^1_N(x, u)=\t V^1_\infty(x).
	\end{equation}
}
\begin{proof}
	By Lemma~\ref{lem:convergence_to_orbit}, for all $x\in\SX$ there exists $\de\in\CL$ such that $\|x_{\t u^1_{N,x}}(N)\|_{\varPi^\star_\SX}\leq\de(N)$. Hence, for $\kappa$ from Assumption~\ref{ass:local_ctrb}, there exists $N_\kappa$ such that $\|x_{\t u^1_{N,x}}(N)\|_{\varPi^\star_\SX}<\kappa$ for $N\geq N_\kappa$.
	Thus, by Assumption~\ref{ass:local_ctrb}, for $N\geq N_\kappa$ we can construct a feasible input $\h u\in\SU^\infty(x)$ by first defining $\h u(k)\coloneq\t u^1_{N,x}(k)$ for $k\in\SI_{[0,N-1]}$, then, for $k\in\SI_{[N,N+M'-1]}$, picking $\h u(k)$ such that $\|x_{\h u}(N+M',x)\|_{\varPi^\star_\SX}=0$ and finally, for $k\geq N+M'$, choosing $\h u(k)$ such that $\|\h u(k)\|_{\varPi^\star_\SU}=0$. For this input it holds that 
	\begin{align*}
		\sum_{k=N}^{\infty}&\t\ell(x_{\h u}(k,x),\h u(k))=
		\sum_{k=N}^{N+M'-1}\t\ell(x_{\h u}(k,x),\h u(k)) \\
		\leq&\sum_{k=N}^{N+M'-1}\bar\al_{\t\ell}\left(\norm{(x_{\h u}(k,x),\h u(k))}_{\varPi^\star}\right) 
		\leq M'\bar\al_{\t\ell}(\rho(\de(N))),
	\end{align*}
	with $\rho\in\CK_\infty$ from Assumption \ref{ass:local_ctrb} and $\bar\al_{\t\ell}\in\CK_\infty$ being the upper bound on $\t\ell$ from \cite[Lemma 27]{Schwenkel2024}. Note that $\de_1\coloneq\bar\al_{\t\ell}\circ\rho\circ\de\in\CL$ and
	\begin{align*}	
		\t V^1_N(x) + M'\de_1(N) 
		\geq \t V^1_N(x) + \sum_{k=N}^{N+M'-1}\t\ell(x_{\h u}(k,x),\h u(k)) \\
		= \t J^1_\infty(x,\h u)
		\geq \inf_{u\in\SU^\infty(x)} \limsup_{T\to\infty} \t J^1_T(x,u),		
	\end{align*}
	where the second inequality follows by optimality. Taking the limit $N\to\infty$ then yields $\t V^1_\infty(x)\geq\min_{u\in\SU^\infty(x)} \limsup_{N\to\infty} \t J^1_N(x,u)$.

	Next, by optimality, $\t J^1_N(x,\t u^1_{\infty,x})\geq\t V^1_N(x)$ for $N\in\N$ and hence $\liminf_{N\to\infty}\t J^1_N(x,\t u^1_{\infty,x})\geq\lim_{N\to\infty}\t V^1_N(x)$.
\end{proof}
%

In the next two lemmas it is shown that no matter whether the input $u^1_{\infty,x}$ or $\t u^1_{\infty,x}$ is applied, the rotated cost $\t J^1_N$ converges as $N\to\infty$. Furthermore, in both cases the optimal state at time $N$ converges to the optimal orbit as $N\to\infty$. 

%
\lemma{ \label{lem:convergence_to_orbit_nonrotated}
	Let Assumptions~\ref{ass:cont_comp}, \ref{ass:strict_dissipativity}, \ref{ass:local_ctrb}, \ref{ass:finite_reachability} and \eqref{eq:infinite_horizon_cost} be satisfied. Then, for all $x\in\SX$ the limit $\lim_{N\to\infty}\t J^1_N(x,u^1_{\infty,x})$ exists and is finite. Moreover, for all $x\in\SX$ there exists $\de\in\CL$ such that $\|x_{u^1_{\infty,x}}(N)\|_{\varPi^\star_\SX}\leq\de(N)$.
}
\begin{proof}
	Since $\bar J^1_N(x,u^1_{\infty,x})$ and $\lb$ are bounded, we infer that also $\t J^1_N(x,u^1_{\infty,x})=\bar J^1_N(x,u^1_\infty)+\lb(x)-\lb(x_{u^1_{\infty,x}}(N))$ is bounded. Furthermore, $\t J^1_N(x,u^1_{\infty,x})$ is monotonically increasing in $N$, i.e., $\t J^1_{N+1}(x,u^1_{\infty,x})-\t J^1_N(x,u^1_{\infty,x})=\t\ell(x_{u^1_{\infty,x}}u^1_{\infty,x}(N))\geq0$.
	
	We can now use similar arguments as in the proof of Lemma~\ref{lem:convergence_to_orbit} after equation \eqref{eq:monotone_increase} to show that for all $x\in\SX$ there exits $\de\in \CL$ satisfying $\|x_{u^1_{\infty,x}}(N)\|_{\varPi^\star_\SX}\leq\de(N)$.
\end{proof}
%
%
\lemma{ \label{lem:convergence_to_orbit_rotated}
	Let Assumptions~\ref{ass:cont_comp}, \ref{ass:strict_dissipativity}, \ref{ass:local_ctrb} and \ref{ass:finite_reachability} be satisfied. Then, for all $x\in\SX$ the limit $\lim_{N\to\infty}\t J^1_N(x,\t u^1_{\infty,x})$ exists and is finite. Moreover, for all $x\in\SX$ there exists $\de\in\CL$ such that $\|x_{\t u^1_{\infty,x}}(N)\|_{\varPi^\star_\SX}\leq\de(N)$.
}
\begin{proof}
	The first statement is an immediate consequence of Lemma~\ref{lem:equiv_rotated_finite_infinte_horizon}. For the second statement we can show a monotonicity property of $\t J^1_N(x,\t u^1_{\infty,x})$ as in the proof of Lemma~\ref{lem:convergence_to_orbit_nonrotated} and then use similar arguments as in the proof of Lemma~\ref{lem:convergence_to_orbit} after equation \eqref{eq:monotone_increase} to show that there exits $\de\in \CL$ satisfying $\|x_{\t u^1_{\infty,x}}(N)\|_{\varPi^\star_\SX}\leq\de(N)$.
\end{proof}
%

Now, Lemma~\ref{lem:convergence_to_orbit_nonrotated} and \eqref{eq:infinite_horizon_cost} can be leveraged to show that the cost along the optimal periodic orbit is zero, which implies that also the storage function is constant along the optimal periodic orbit.
%
\lemma{ \label{lem:storage_on_orbit}
	Let Assumptions~\ref{ass:cont_comp}, \ref{ass:strict_dissipativity}, \ref{ass:local_ctrb}, \ref{ass:finite_reachability} and \eqref{eq:infinite_horizon_cost} be satisfied. Then, $\lb(\varPi^\star_\SX(i))=\lb(\varPi^\star_\SX(j))$ for all $i,j\in\SI_{[0,p^\star-1]}$.
}
\begin{proof}
	By \eqref{eq:infinite_horizon_cost} we know that $\lim_{N\to\infty}\sum_{k=0}^{N-1}\bar\ell(x_{u^1_{\infty,x}}(k),u^1_{\infty,x}(k))$ exists and hence
	\begin{align}
		\lim_{k\to\infty}\bar\ell(x_{u^1_{\infty,x}}(k),u^1_{\infty,x}(k))=0. \label{eq:cost_on_opt_orbit}
	\end{align}
	Also, by Lemma~\ref{lem:convergence_to_orbit_nonrotated}, $\lim_{N\to\infty}\sum_{k=0}^{N-1}\t\ell(x_{u^1_{\infty,x}}(k),u^1_{\infty,x}(k))$ exists and hence
	\begin{align}
		\nonumber\quad0&=\lim_{k\to\infty}\t\ell(x_{u^1_{\infty,x}}(k),u^1_{\infty,x}(k)) \\
		\nonumber&\geq\lim_{k\to\infty}\ubar{\al}_{\t\ell}(\|(x_{u^1_{\infty,x}}(k),u^1_{\infty,x}(k))\|_{\varPi^\star})\geq0,
	\end{align}
	which yields $0=\lim_{k\to\infty}\|(x_{u^1_{\infty,x}}(k),u^1_{\infty,x}(k))\|_{\varPi^\star}$.
	Combining this with \eqref{eq:cost_on_opt_orbit} implies due to continuity of $\bar\ell$ that $\bar\ell(\varPi^\star(k))=0$ for all $k\in\SI_{[0,p^\star-1]}$, which in turn implies that $\lb(\varPi^\star_\SX(i))=\lb(\varPi^\star_\SX(j))$ for all $i,j\in\SI_{[0,p^\star-1]}$.
\end{proof}
%

The fact that the storage function $\lb$ is constant along the optimal orbit can now be used to show that the non-rotated optimal infinite-horizon input $u^1_{\infty,x}$ is a minimizing input for the rotated infinite-horizon cost $\t J^1_\infty$.

%
\lemma{ \label{lem:same_minimizer}
	Let Assumptions~\ref{ass:cont_comp}, \ref{ass:strict_dissipativity}, \ref{ass:local_ctrb}, \ref{ass:finite_reachability} and \eqref{eq:infinite_horizon_cost} be satisfied. Then,
	$\t J^1_\infty(x,u^1_{\infty,x})=\t J^1_\infty(x,\t u^1_{\infty,x})$ holds for all $x\in\SX$.
}
\begin{proof}
	For any $x\in\SX$ a quick calculation yields
	\begin{equation*}
		\t J^1_N(x,u^1_{\infty,x}) - \bar J^1_N(x,u^1_{\infty,x}) = \lb(x) - \lb(x_{u^1_{\infty,x}}(N))
	\end{equation*}
	and by Lemma~\ref{lem:convergence_to_orbit_nonrotated} there exists $\de\in\CL$ such that $\|x_{u^1_{\infty,x}}(N)\|_{\varPi_\SX}\leq\de(N)$. Thus, we can use continuity of $\lb$, Lemma~\ref{lem:continuity} in the Appendix and Lemma~\ref{lem:storage_on_orbit} to infer
	\begin{equation} \label{eq:eq1}
		\abs{\t J^1_N(x,u^1_{\infty,x}) - \bar J^1_N(x,u^1_{\infty,x}) -\lb(x) + \lb_{\varPi^\star}} \leq \al_\lb(\de(N)),
	\end{equation}
	with $\al_\lb\in\CK_\infty$. By the same arguments (we only have to replace Lemma~\ref{lem:convergence_to_orbit_nonrotated} by Lemma~\ref{lem:convergence_to_orbit_rotated}) we obtain the existence of $\t\de\in\CL$ such that
	\begin{equation} \label{eq:eq2} 
		\abs{\t J^1_N(x,\t u^1_{\infty,x}) - \bar J^1_N(x,\t u^1_{\infty,x}) -\lb(x) + \lb_{\varPi^\star}} \leq \al_\lb(\t \de(N)).
	\end{equation}
	Note that the limit $N\rightarrow\infty$ of each individual term in the expression above exists. Taking the limit $N\to\infty$ and combining \eqref{eq:eq1} and \eqref{eq:eq2} then yields 
	\begin{equation*}
		\t J^1_\infty(x,u^1_{\infty,x})+\bar J^1_\infty(x,\t u^1_{\infty,x}) = \t J^1_\infty(x,\t u^1_{\infty,x})+\bar J^1_\infty(x,u^1_{\infty,x})
	\end{equation*}
	and by optimality also
	\begin{equation*}
		\t J^1_\infty(x,u^1_{\infty,x})+\bar J^1_\infty(x, u^1_{\infty,x}) \leq \t J^1_\infty(x,\t u^1_{\infty,x})+\bar J^1_\infty(x,u^1_{\infty,x}),
	\end{equation*}
	which implies $\t J^1_\infty(x,u^1_{\infty,x})\leq\t J^1_\infty(x,\t u^1_{\infty,x})$.  By optimality we also have $\t J^1_\infty(x,u^1_{\infty,x})\geq\t J^1_\infty(x,\t u^1_{\infty,x})$ and hence $\t J^1_\infty(x,u^1_\infty)=\t J^1_\infty(x,\t u^1_{\infty,x})$.
\end{proof}
%

We are now finally ready to prove the second part of Theorem~\ref{lem:difference_rotated_value_fcns}.

\begin{proof}
	By Lemma~\ref{lem:same_minimizer}, \eqref{eq:eq1} and \eqref{eq:infinite_horizon_cost} we infer that
	\begin{align*}
		\t J^1_\infty(x,\t u^1_{\infty,x})=\t J^1_\infty(x,u^1_{\infty,x}) = \bar V_\infty(x)-\lb(x)+\lb_{\varPi^\star}
	\end{align*}
	and by Lemma~\ref{lem:equiv_rotated_finite_infinte_horizon} that $\t J^1_\infty(x,\t u^1_{\infty,x})=\t V^1_\infty(x)$. 
	Also, in the proof of the first part of Theorem~\ref{lem:difference_rotated_value_fcns} it was shown that $\t V^1_\infty(x)=\t V^\mathrm{ces}_\infty(x)$. Hence, by Lemma~\ref{lem:convergence_value_fcn} and the relations above, we have 
	\begin{align*}
		\bar V^\mathrm{ces}_\infty(x)-\lb(x)+\lb_{\varPi^\star}=\t V^\mathrm{ces}_\infty(x)	= \bar V_\infty(x)-\lb(x)+\lb_{\varPi^\star},
	\end{align*}
which proves $\bar V_\infty(x)=\bar V^\mathrm{ces}_\infty(x)$.
\end{proof}

\section{Finite State Setting} \label{sec:finite_states}

In this section, we consider a finite state set as a special case, which is a setting, where the VI can be numerically solved. In particular, we show that such a setting is always optimally operated at a periodic orbit and we show that the Assumptions \ref{ass:strict_dissipativity} and \ref{ass:local_ctrb} are satisfied if there is a unique minimal optimal periodic orbit.

\ass[Finite state set]{\label{ass:finiteStateSpace}
	The state set $\SX\subset\R^n$ satisfies $N_\SX\coloneq\#\SX <\infty$.
}

In the finite state setting, we can remove from any trajectory periodic orbits of length $p\leq N_\SX$ until we are left with a transient part of length less than $N_\SX$. 
This is exploited in the following Lemma.
%
\lemma{ \label{lem:trajectory_decomposition}
	Let Assumptions \ref{ass:cont_comp} and \ref{ass:finiteStateSpace} be satisfied, $x\in\SX$, $g:\SX\times\SU\rightarrow\R$ and let $\#{\varPi}$ denote the period length of an orbit $\varPi$. For any $N\in\N$ and any input trajectory $u\in\SU^N(x)$, there exist $N_\varPi\in\N$, a sequence of periodic orbits $(\varPi_i)_{i\in\SI_{[0,N_\varPi]}}$, $\varPi_i\in S^p_\varPi$ with $1\leq p\leq N_\SX$ and a set $R(x,u)\subseteq\SI_{[0,N-1]}$ with $\#R(x,u) < N_\SX$ such that
	\begin{align} \label{eq:trajectory_decomposition}
		\begin{split}
			\sum_{k=0}^{N-1}&g(x_u(k,x),u(k)) \\
			=&\sum_{i=0}^{N_\varPi}\sum_{j=0}^{\#{\varPi_i}-1}g(\varPi_i(j))+\sum_{j\in R(x,u)}g(x_u(j,x),u(j)).
		\end{split}
	\end{align}
}
\begin{proof}
	We first show that for all $k_0\in [0,N-N_\SX+1]$ and $u \in U^N(x)$, the trajectory $(x_u(k,x),u(k))$, $k\in\SI_{[k_0,k_0+N_\SX-1]}$ of length $N_\SX$ contains at least one orbit $\varPi\in S^p_\varPi$ with $p\leq N_\SX$. Suppose this was not the case. Then there exists $u\in\SU^{N_\SX}(x)$ satisfying $x_u(j,x)\neq x_u(k,x)$ for all $j\neq k$, $j,k\in\SI_{[k_0,k_0+N_\SX]}$. This is a contradiction, since $x_u(k,x)\in\SX$ for all $k\in\SI_{[0,N]}$ and $\#\SX=N_\SX$.
	
	Next, we construct the sequence of orbits $\varPi_i$ by repeatedly performing the following steps, starting with $N_0\coloneq N$ and $u_0\coloneq u$. Consider the state-input trajectory $(x_{u_i}(k,x),u_i(k))$, $k\in\SI_{[0,N_i-1]}$ with $N_i\geq N_\SX$. By the argument above, there exist $k_1,k_2\in\SI_{[0,N_i-1]}$, $0\leq k_2-k_1<N_\SX$ such that
	$\varPi_i \coloneq \big( (x_{u_i}(k,x), u_i(k)) \big)_{k\in \SI_{[k_1,k_2]}}$ is an orbit $\varPi_i\in S^{p_i}_\varPi$ with $p_i\coloneq k_2-k_1+1$. Next, remove $\varPi_i$ from the trajectory, i.e., define $u_{i+1}(k)\coloneq u_i(k)$, $k\in\SI_{[0,k_1-1]}$ and $ u_{i+1}(k)\coloneq u_i(k+p_i)$, $k\in\SI_{[k_1,N_i-p_i]}$ to obtain $(x_{u_{i+1}}(k,x),u_{i+1}(k))$, $k\in\SI_{[0,N_{i+1}-1]}$ with $N_{i+1}\coloneq N_i-p_i$. Since we removed an orbit, this is a feasible trajectory and hence we can repeat the steps above. If $N_{i+1}<N_\SX$, then the $\varPi_i$ form the desired sequence and the existence of the set $R(x,u)$ is obvious.
\end{proof}

Even more, the existence of an optimal periodic orbit is already guaranteed in the finite state setting.

\lemma{ \label{lem:existence_orbit}
	Let Assumptions \ref{ass:cont_comp} and \ref{ass:finiteStateSpace} be satisfied. Then, there exists an optimal periodic orbit $\varPi^\star\in S^{p^\star}_\varPi$ according to \eqref{eq:opt_orbit}, which has length $p^\star\leq N_\SX$.
}
\begin{proof}
	Lemma \ref{lem:trajectory_decomposition} implies that there exist periodic orbits $\varPi\in S^p_\varPi$ with $p\leq N_\SX$ and that every orbit $\varPi\in S^{\h p}_\varPi$ of length $\h p>N_\SX$ can be decomposed in shorter orbits of length $p\leq N_\SX$. Therefore, we only need to consider periodic orbits $\varPi\in S^p_\varPi$ with $p\leq N_\SX$. Since $N_\SX<\infty$, there only exist finitely many such orbits and therefore there exists at least one optimal periodic orbit $\varPi^\star\in S^{p^\star}_\varPi$ with $p^\star\leq N_\SX$.
\end{proof}

Rather weak additional assumptions are already sufficient to prove strict dissipativity w.r.t. $\varPi^\star$. 

\ass[$\varPi^\star$ is minimal \& unique]{\label{ass:minOrbit}
The optimal periodic orbit $\varPi^\star$ from Lemma \ref{lem:existence_orbit} is minimal. Further, if $\ell^p(\varPi) = \ell^{p^\star}(\varPi^\star)$, then $\|\varPi(k)\|_{\varPi^\star} = 0$ for all $k$ in $[0,p-1]$.
}

\theorem[Strict dissipativity]{
	If Assumption \ref{ass:finiteStateSpace} is satisfied, then also Assumption \ref{ass:local_ctrb} is satisfied. If additionally Assumptions \ref{ass:cont_comp}, \ref{ass:finite_reachability}, \ref{ass:minOrbit} hold, then Assumption \ref{ass:strict_dissipativity} is satisfied.
}
\begin{proof}
	By Assumption \ref{ass:finiteStateSpace}, Assumption \ref{ass:local_ctrb} is trivially satisfied for the choice $0<\kappa<\tfrac{1}{2}\min_{x,y\in\SX,x\neq y}\norm{x-y}$, since then, for $x\in\SX$, $z\in\varPi^\star_\SX$, $\norm{x-z}\leq\kappa$ implies $x\in\varPi^\star_\SX$.
	For the second statement, since there only exist finitely many different orbits of length $p\leq N_\SX$, all non-optimal periodic orbits have an average cost strictly larger than $\ell^{p^\star}(\varPi^\star)$. Hence, we can define
	\begin{gather*}
		\de\coloneq\min_{\substack{\varPi\in S_\varPi^p \\ p\in\SI_{[1,N_\SX]}\\\ell^p(\varPi)\neq\ell^{p^\star}(\varPi^\star)}} \ell^p(\varPi)-\ell^{p^\star}(\varPi^\star)>0,
	\end{gather*}
	Based on that definition of $\de$, we choose $\al\in\CK_\infty$ such that 
	\begin{gather*}
		\al_{\max}\coloneq\max_{(x,u)\in\SX\times\SU}\al(\norm{(x,u)}_{\varPi^\star})<\de.
	\end{gather*}
	Thus, summing up the supply rate $s(x,u)\coloneq\ell(x,u)-\ell^{p^\star}(\varPi^\star)-\al(\norm{(x,u)}_{\varPi^\star})$ over an orbit $\varPi\in S_\varPi^p$ yields
	\begin{gather}\label{eq:supply_of_orbit}
		\sum_{i=0}^{p-1}\bigl(\ell(\varPi(i))-\ell^{p^\star}(\varPi^\star)-\al(\norm{\varPi(i)}_{\varPi^\star})\bigr)\geq 0.
	\end{gather}
	This sum is non-negative, as due to the definition of $\de$ and $\al$, it can only be negative if $\ell^p(\varPi) = \ell^{p^\star}(\varPi^\star)$. However, if $\ell^p(\varPi) = \ell^{p^\star}(\varPi^\star)$, then $\|\varPi(i)\|_{\varPi^\star} =0$ for all $i\in[0,p-1]$ due to Assumption~\ref{ass:minOrbit}.
	It is well-known, that a system is dissipative with respect to the supply rate $s(x,u)$ and with a bounded storage function $\lb:\SX\to\R$ if the available storage
	\begin{gather*}
		S_\mathrm{a}(x)\coloneq\sup_{N\in\N,u\in\SU^N(x)}\sum_{k=0}^{N-1} -s(x_u(k,x),u(k))
	\end{gather*}
	is bounded on $\SX\vphantom{\h\ell}$ \cite{Willems1972}\footnote{While \cite{Willems1972} considered continuous-time systems without constraints, \cite{Mueller2015} argued that these results can be analogously obtained in the discrete time case with state and input constraints.}. By combining \eqref{eq:supply_of_orbit} with Lemma \ref{lem:trajectory_decomposition} we infer that $S_\mathrm{a}(x) \leq N_\SX(\al_{\max}-\ell_{\min}+\ell^{p^\star}(\varPi^\star))$ for all $x\in\SX$. Lemma \ref{lem:trajectory_decomposition}, Assumption \ref{ass:finite_reachability} and \eqref{eq:supply_of_orbit} also imply for all $x\in\SX$ that $S_\mathrm{a}(x) \geq -N_\SX(\ell_{\max}-\ell^{p^\star}(\varPi^\star))$, since the system can be steered to the optimal periodic orbit and remain there. Hence, we have dissipativity w.r.t. the supply rate $s(x,u)$, which implies $\ell^{p^\star}(\varPi^\star)=\ell^\star$ as discussed after assumption \ref{ass:strict_dissipativity} and hence Assumption  \ref{ass:strict_dissipativity} is satisfied (Technically, we have not shown continuity of $\lb$, but since $\#\SX= N_\SX$, we can continuously interpolate $\lb$ to obtain a continuous storage function).
\end{proof}

Hence, all guarantees derived in Section \ref{sec:convergence} are also valid under Assumptions \ref{ass:cont_comp}, \ref{ass:finite_reachability}, \ref{ass:finiteStateSpace}, and \ref{ass:minOrbit}. Further, the guarantees of Section \ref{sec:discounts} are valid, if additionally Assumption \ref{ass:discount_fcn} holds.


\section{Numerical Example} \label{sec:numerics}

\begin{figure}[tp]
	\medskip
	\begin{tikzpicture}
		\begin{axis} [grid=major, legend pos=south east, legend style={font=\tiny,  legend columns=-1},
			ymin=-1.6, ymax=2.44,
			xmin=0, xmax=40, xlabel={$N$},
			height=4cm, width = \columnwidth,
			cycle list name=convPlot]
			\addplot table [x index=0, y index=6, col sep=comma] {ces_value_fcn.txt};
			\addlegendentry{$\bar V^\mathrm{ces}_N(6)$}
			\addplot table [x index=0, y index=6, col sep=comma] {value_fcn08.txt};
			\addlegendentry{$\bar V^{0.8}_N(6)$}
			\addplot table [x index=0, y index=6, col sep=comma] {value_fcn06.txt};
			\addlegendentry{$\bar V^{0.6}_N(6)$}
		\end{axis}
	\end{tikzpicture}
	\caption{Values of value functions for $x=6$ in dependence of $N$.}
	\label{Fig:value_fcn}
\end{figure}
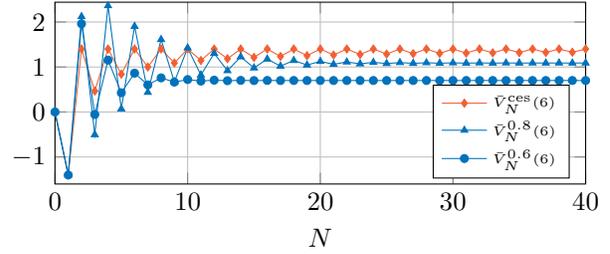

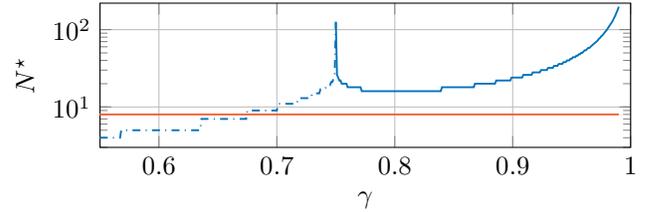
\begin{figure}
	\begin{tikzpicture}
		\begin{axis} [grid=major, 
			ymin=3, ymax=220, ymode=log, ylabel={$N^\star$},
			xmin=0.55, xmax=1, xlabel={$\ga$},
			height=3.5cm, width = \columnwidth,
			cycle list name=stepsPlot]
			\addplot table [x index=0, y index=1, col sep=comma, mark=none] {steps1.txt}; \label{plt:steps1}
			\addplot table [x index=0, y index=1, col sep=comma, mark=none] {steps2.txt}; \label{plt:steps2}
			\addplot table [x index=0, y index=2, col sep=comma, mark=none] {steps.txt}; \label{plt:steps}
		\end{axis}
	\end{tikzpicture}
	\caption{Number of VIs $N^\star$ of $\bar V^\mathrm{ces}_N$ (red) and $\bar V^\ga_N$ (blue) that are required for the resulting policy to converge. A dashed-dotted line indicates that the resulting input is sub-optimal w.r.t. Cesàro infinite-horizon performance.}
	\label{Fig:steps}
\end{figure}

In this section, for a linear system and a finite state set, we compare the CVI to a VI with constant discount factor $\ga\in(0,1)$, i.e., for all $x\in\SX$,
\begin{gather*}
	\bar V^\ga_{N+1}(x) = \min_{u\in\SU^1(x)} \bar\ell(x,u) + \ga \bar V^\ga_{N}(f(x,u)), \quad \bar V^\ga_0(x)=0.
\end{gather*}
Consider the linear system with state and input constraints
\begin{gather*}
	x(k+1)=x(k)+u(k), \quad
	\quad x\in\SI_{[1,9]}, \quad u\in\SI_{[-8,8]},
\end{gather*}
and the stage cost $\ell(x,u)=0.1\left((x-4)^2-(u-4)^2\right)+6.3$. This system is optimally operated at the unique and minimal optimal periodic orbit $\varPi^\star=\left((1,8),(9,-8)\right)$ with $p^\star=2$ and $\ell^\star=0$, i.e., $\bar\ell=\ell$.
Both $\bar V^\mathrm{ces}_N$ and $\bar V^\ga_N$ converge with increasing $N$. However, the value of convergence of $V^\ga_N$ depends on $\ga$ and the limit is therefore not equal to $\bar V^\mathrm{ces}_\infty$ as can be seen in Fig \ref{Fig:value_fcn} for $x=6$. Because of this gap, also the resulting optimal policies might differ. For example, starting at $x=4$, the CVI suggests to first go to $x=9$ and then to $x=1$, with combined cost of $\bar\ell(4,5)+\bar\ell(9,-8)=0.6$, while the VI of $\bar V^{0.6}_N$ suggests to directly go to $x=1$ with a cost of $\bar\ell(4,-3)=1.4$. Hence, a VI with a fixed discount $\ga$ might lead to a sub-optimal policy in the Cesàro infinite-horizon sense. Moreover, Fig. \ref{Fig:steps} illustrates how many iterations are required until the resulting policy converged in dependence of $\ga$ and in comparison to the CVI. As it can be seen, when both VIs lead to the same optimal policy, then the CVI needs less iterations, or when $\bar V_N^\ga$ is faster, then it provides a sub-optimal policy.

\pgfplotscreateplotcyclelist{convPlot}{RedOrange,line width=0.4pt,mark size=1.5pt,mark=diamond*\\%
	RoyalBlue,line width=0.4pt,mark size=1.5pt,mark=triangle*\\%
	RoyalBlue,line width=0.4pt,mark size=1.5pt,mark=*\\}

\pgfplotscreateplotcyclelist{stepsPlot}{dashdotted,RoyalBlue,line width=0.7pt\\solid,RoyalBlue,line width=0.7pt\\ solid,RedOrange,line width=0.7pt\\}

\section{Conclusion} \label{sec:conclusion}

In this work, we used the Cesàro mean to define the infinite-horizon optimal control problem in cases where the undiscounted VI does not converge to the optimal infinite-horizon cost. Furthermore, by introducing the CVI, we provided a recursive procedure to numerically calculate the Cesàro value function, and proved its convergence for systems that are optimally operated at a periodic orbit. We further showed that the Cesàro value function is consistent with the infinite horizon undiscounted cost, if the latter limit exists and is finite.

Our work opens up several future research questions. First of all, how can the CVI in practice be carried out for an uncountable infinite state and input space? Next, can convergence of the CVI be guaranteed for even larger problem classes? This could potentially include optimal operation at non-minimal or quasi-periodic orbits. Furthermore, in the deterministic case with uncountable infinite state and input spaces, can the factor $\tfrac{N-1}{N}$ in the Cesàro value iteration be replaced with a more general class of factors $\alpha_N\to1$ as $N\to\infty$ as was done in \cite{hordijk1975modified} in the finite state setting for MDPs? Lastly, it would also be interesting to analyze the CVI in detail on more complex examples.

\bibliographystyle{IEEEtran}
\bibliography{../../../literature/Literature}

\appendix

\lemma{\label{lem:uniform_bound}
	Let Assumptions \ref{ass:cont_comp}, \ref{ass:strict_dissipativity}, \ref{ass:local_ctrb}, \ref{ass:finite_reachability} and \ref{ass:discount_fcn} hold. Then, there exists $\t C<\infty$ such that $\t V^\be_N(x)\leq \t C$ for all $N\in\N$ and for all $x\in\SX$.
}
\begin{proof}
	By Assumptions \ref{ass:local_ctrb} and \ref{ass:finite_reachability}, for all $x\in\SX$, there exist $\h u\in\SU^M(x)$ with $M\coloneq M'+M''<\infty$ and $l\in\SI_{[0,p^\star-1]}$ such that $x_{\h{u}}(M,x)=\varPi^\star_\SX(l)$.
	For $N\geq M$, we extend this input $\h{u}$ to any horizon length $N$ by choosing $\h{u}(k)\coloneq\varPi^\star_\SU([k-M+l]_{p^\star})$ for $k\in\SI_{[M,N-1]}$. The corresponding state then satisfies $x_{\h{u}}(k,x)=\varPi^\star_\SX([k-M+l]_{p^\star})$ for all $k\in\SI_{[M,N-1]}$. Note that $\t\ell(\varPi^\star([j]_{p^\star}))=0$ for all $j\in\N$ (see \cite[Lemma 27]{Schwenkel2024}). Therefore, 
	\begin{alignat*}{2}
		\t V_N^\be(x)\leq& \t J^\be_N(x,\h{u})
		=\!\sum_{k=0}^{M-1}\be(\tfrac{k}{N})\t\ell(x_{\h{u}}(k,x),\h{u}(k)) 
		\leq M\t\ell_{\max}.
	\end{alignat*}
	\\[-0.6cm]
\end{proof}

\lemma{ \label{lem:monotonicity_rot_value_fcn}
	Let Assumption \ref{ass:discount_fcn} hold. Then, the rotated value function is monotonically increasing w.r.t. $N$, i.e. $\t V^\be_{N+1}(x)\geq\t V^\be_N(x)$ for all $x\in\SX$ and $N\in\N$.}
\begin{proof}
	With the optimal input $\t u^\be_{N+1,x}$ we infer
	\begin{gather*}
		\t V^\be_{N+1}(x)-\t V^\be_N(x) \geq \t J^\be_{N+1}(x,\t u^\be_{N+1,x}) - \t J^\be_{N}(x,\t u^\be_{N+1,x}) \\
		= \sum_{k=0}^{N}(\be(\tfrac{k}{N+1})-\be(\tfrac{k}{N}))\t\ell(x_{\t u^\be_{N+1,x}}(k,x),\t u^\be_{N+1,x}(k))	\geq 0,
	\end{gather*}
	where the last inequality follows from $\t \ell(x,u)\geq0$ for all $(x,u)\in\SX\times\SU$, $\be(1)=0$ and $\be(\tfrac{k}{N+1})-\be(\tfrac{k}{N})\geq0$, since $\be$ is non-increasing and $\tfrac{k}{N+1}<\tfrac{k}{N}$.
\end{proof}

\lemma{ \label{lem:rotated_cost_fcn}
	Let Assumption \ref{ass:discount_fcn} hold. Then, for all $N\in\N$, $x\in\SX$ and $u\in\SU^N(x)$, the rotated cost function satisfies
	\begin{align*}
		\t J^\be_N(x,u)=&~\bar J^\be_N(x,u)+\lb(x) \\
		&-\sum_{k=1}^{N}(\be(\tfrac{k-1}{N})-\be(\tfrac{k}{N}))\lb(x_u(k,x)).
	\end{align*}
}
\begin{proof}
	A simple calculation shows that $\t J^\be_N(x,u)=$
	\begin{align*}
		\bar J^\be_N(x,u)
		+\sum_{k=0}^{N-1}\be(\tfrac{k}{N})\big(\lb(x_u(k,x))-\lb(x_u(k+1,x))\big).
	\end{align*}
	Using $\be(0)=1$ and $\be(1)=0$ yields
	\begin{align*}
		\sum_{k=0}^{N-1}\be&(\tfrac{k}{N})(\lb(x_u(k,x))-\lb(x_u(k+1,x))) \\
		=& \be(0)\lb(x) - \be(\tfrac{N-1}{N})\lb(x_u(N,x)) \\
		&+ \sum_{k=1}^{N-1}(\be(\tfrac{k}{N})-\be(\tfrac{k-1}{N}))\lb(x_u(k,x)) \\
		=& \lb(x) - \sum_{k=1}^{N}(\be(\tfrac{k-1}{N})-\be(\tfrac{k}{N}))\lb(x_u(k,x)).
	\end{align*}
	\\[-0.8cm]
\end{proof}
	\lemma{\label{lem:continuity}
	For any continuous function $g$ defined on a compact set $\SY$ there exists $\al_g\in\CK_\infty$ such that $\norm{g(y)-g(y')}\leq\al_g(\norm{y-y'})$ for all $y,y'\in\SY$.
}
\lemma{ \label{lem:follow_orbit}
	Let Assumption~\ref{ass:cont_comp} be satisfied and $\eps>0$. There exists $\al_f\in\CK_\infty$ such that $\|f(x,u)\|_{\varPi_\SX}<\al_f(\eps)$ for all $(x,u)\in\SX\times\SU$ satisfying $\|(x,u)\|_{\varPi^\star}<\eps$.
}
\begin{proof}
	Consider $l\in\SI_{[0,p^\star-1]}$ such that $\|(x,u)\|_{\varPi^\star}=\|(x,u)-\varPi^\star(l)\|$. By continuity of $f$ and Lemma~\ref{lem:continuity}, there exists $\al_f\in\CK_\infty$ such that for all $(x,u)\in\SX\times\SU$ we have
	\begin{align*}
		\begin{split}
			\al_f&(\norm{(x,u)}_{\varPi^\star})=\al_f(\norm{(x,u)-\varPi^\star(l)})\\
			&\geq\norm{f(x,u)-\varPi^\star_\SX([l+1]_{p^\star})}\geq	\norm{(f(x,u))}_{\varPi^\star_\SX}.
		\end{split}
	\end{align*}
	Exploiting $\|(x,u)\|_{\varPi^\star}<\eps$ then yields the claim.
\end{proof}

\end{document}